\documentclass[iop,apj,tighten]{emulateapj}
\newcommand\submitms{n}		% set to y to follow AAS ``ms'' names, etc.
\if\submitms y
\else
\usepackage{apjfonts}
\fi
\usepackage{ifthen}
\usepackage{natbib}

\shorttitle{THE 1998 NOVEMBER 14 SATURN OCCULTATION. II.}
\shortauthors{HARRINGTON, FRENCH, \& MATCHEVA}

\newcommand\phoz{\phantom{{\pm} 0.0}}
\newcommand\phtz{\phantom{{\pm} 0.00}}
\newcommand\phrz{\phantom{{\pm} 0.000}}
\newcommand\phfz{\phantom{{\pm} 0.0000}}
\newcommand\degree{\degr}
\newcommand\degrees\degree
\newcommand\decdegree{\fdg}
\newcounter{fignum}

\DeclareSymbolFont{UPM}{U}{eur}{m}{n}
\DeclareMathSymbol{\umu}{0}{UPM}{"16}
\let\oldumu=\umu
\renewcommand\umu{\ifmmode\oldumu\else\math{\oldumu}\fi}
\newcommand\micro{\umu}
\renewcommand\micro{\math{\mu}}%ApJ only
\renewcommand\micron{\micro m}
\newcommand\microns \micron
\newcommand\microbar{\micro bar}

\let\oldsim=\sim
\renewcommand\sim{\ifmmode\oldsim\else\math{\oldsim}\fi}
\let\oldpm=\pm
\renewcommand\pm{\ifmmode\oldpm\else\math{\oldpm}\fi}
\newcommand\by{\ifmmode\times\else\math{\times}\fi}
\newcommand\ttt[1]{10\sp{#1}}
\newcommand\tttt[1]{\by\ttt{#1}}
\newcommand\tablebox[1]{\begin{tabular}[t]{@{}l@{}}#1\end{tabular}}
\newbox{\wdbox}
\renewcommand\c{\setbox\wdbox=\hbox{,}\hspace{\wd\wdbox}}
\renewcommand\i{\setbox\wdbox=\hbox{i}\hspace{\wd\wdbox}}

\newcommand\herenote[1]{{\bfseries #1}\typeout{======================> note on page \arabic{page} <====================}}

\newcount\timect
\newcount\hourct
\newcount\minct
\newcommand\now{\timect=\time \divide\timect by 60
         \hourct=\timect \multiply\hourct by 60
         \minct=\time \advance\minct by -\hourct
         \number\timect:\ifnum \minct < 10 0\fi\number\minct}

\catcode`@=11

\if\submitms y
\renewcommand\comment[1]{}
\else
\newcommand\comment[1]{}
\fi
\newcommand\commenton{\catcode`\%=14}
\newcommand\commentoff{\catcode`\%=12}

\renewcommand\math[1]{$#1$}
\newcommand\mathshifton{\catcode`\$=3}
\newcommand\mathshiftoff{\catcode`\$=12}

\comment{the backslash is necessary}

\comment{alignment tab}

\let\atab=&
\newcommand\atabon{\catcode`\&=4}
\newcommand\ataboff{\catcode`\&=12}

\let\oldmsp=\sp
\let\oldmsb=\sb
\renewcommand\sp[1]{\ifmmode
	   \oldmsp{#1}%
	 \else\strut\raise.85ex\hbox{\scriptsize #1}\fi}
\renewcommand\sb[1]{\ifmmode
	   \oldmsb{#1}%
	 \else\strut\raise-.54ex\hbox{\scriptsize #1}\fi}
\newcommand\msp[1]{\ifmmode
	   \oldmsp{#1}
	 \else \math{\oldmsp{#1}}\fi}
\newcommand\msb[1]{\ifmmode
	   \oldmsb{#1}
	 \else \math{\oldmsb{#1}}\fi}
\newcommand\supon{\catcode`\^=7}
\newcommand\supoff{\catcode`\^=12}
\newcommand\subon{\catcode`\_=8}
\newcommand\suboff{\catcode`\_=12}
\newcommand\supsubon{\supon \subon}
\newcommand\supsuboff{\supoff \suboff}

\newcommand\actcharon{\catcode`\~=13}
\newcommand\actcharoff{\catcode`\~=12}

\newcommand\paramon{\catcode`\#=6}
\newcommand\paramoff{\catcode`\#=12}

\comment{And now to turn us totally on and off...}

\newcommand\reservedcharson{\commenton \mathshifton \atabon \supsubon \actcharon
	\paramon}

\newcommand\reservedcharsoff{\commentoff \mathshiftoff \ataboff
	\supsuboff \actcharoff \paramoff}

\newcommand\nojoe[1]{\reservedcharson#1\reservedcharsoff}

\catcode`@=12
\reservedcharsoff

\reservedcharson

\comment{ Must have ONLY ONE of these... trust these macros, they work

}
\reservedcharsoff

\actcharon
\if\submitms y
\newcommand\widedeltab{deluxetable}
\else
\newcommand\widedeltab{deluxetable*}
\received{2009 July 23}
\comment{\revised{2010 February 21}}
\accepted{2010 April 7}
\published{2010 May 19}
\fi

\if\submitms y
\slugcomment{\tablebox{
\comment{Submitted to {\em ApJ} Mon Oct  4 21:52:00 EDT 2004.}
Submitted to {\em ApJ} 2009 July 23.\\
\comment{Revised 2010 February 21.\\}
Accepted 2010 April 7.}}
\else
\slugcomment{Version of 2010 May 19 for arXiv.org.}
\journalinfo{{\em To appear in} The Astrophysical Journal, 716:404--416, 2010 June 10.
             \hfill  {\em as} {\rm doi}:10.1088/0004-637X/716/1/404}
\fi

\begin{document}

\title{The 1998 November 14 Occultation of GSC 0622-00345 by Saturn.\\
  II. Stratospheric Thermal Profile, Power Spectrum, and Gravity Waves}
\author{Joseph Harrington\altaffilmark{1,4}}
\author{Richard G.\ French\altaffilmark{2,4}}
\author{Katia Matcheva\altaffilmark{3}}
\affil{\sp{1}Planetary Sciences Group, Department of Physics, University of
  Central Florida, Orlando, FL 32816-2385, USA; jh@physics.ucf.edu}
\affil{\sp{2}Astronomy Department, Wellesley College, Wellesley, MA
  02481, USA; rfrench@wellesley.edu}
\affil{\sp{3}Department of Physics, University of Florida, P.O.\ Box 118440,
  Gainesville, FL 32611, USA; katia@phys.ufl.edu}
\altaffiltext{4}{Visiting Astronomer at the Infrared Telescope
  Facility, which is operated by the University of Hawaii under
  Cooperative Agreement no.\ NCC 5-538 with the National Aeronautics
  and Space Administration, Science Mission Directorate, Planetary
  Astronomy Program.}

\begin{abstract}

On 1998 November 14, Saturn and its rings occulted the star GSC
0622-00345.  The occultation latitude was 55{\decdegree}5 S.  This paper
analyzes the \mbox{2.3 {\micron}} light curve derived by
\citeauthor{HarringtonFrench2010apjsatoc98I}.  A
fixed-baseline isothermal fit to the light curve has a temperature of
\mbox{140 {\pm} 3 K}, assuming a mean molecular mass of 2.35 AMU.  The
thermal profile obtained by numerical inversion is valid between 1 and
60 {\micro}bar.  The vertical temperature gradient is \math{>}0.2 K
km\sp{-1} more stable than the adiabatic lapse rate, but it still
shows the alternating-rounded-spiked features seen in many temperature
gradient profiles from other atmospheric occultations and usually
attributed to breaking gravity (buoyancy) waves.  We conduct a wavelet
analysis of the thermal profile, and show that, even with our low
level of noise, scintillation due to turbulence in Earth's
atmosphere can produce large temperature swings in light-curve
inversions.  Spurious periodic features in the ``reliable'' region of
a wavelet amplitude spectrum can exceed 0.3 K in our data.  We also
show that gravity-wave model fits to noisy isothermal light curves can
lead to convincing wave ``detections''.  We provide new significance
tests for localized wavelet amplitudes, wave model fits, and global
power spectra of inverted occultation light curves by assessing the
effects of pre- and post-occultation noise on these parameters.  Based
on these tests, we detect several significant ridges and isolated
peaks in wavelet amplitude, to which we fit a gravity wave model.  We
also strongly detect the global power spectrum of thermal fluctuations
in Saturn's atmosphere, which resembles the ``universal'' (modified
Desaubies) curve associated with saturated spectra of propagating
gravity waves on Earth and Jupiter.

\if\submitms y
\else
\comment{\hfill\herenote{DRAFT of {\today} \now}.}
\fi
\end{abstract}

\section{INTRODUCTION}
\label{intro}

Earth-based occultations remain an attractive method for measuring the
thermal profile in the 1--100-{\micro}bar region of a planetary
atmosphere.  Many such profiles for Saturn were recorded during the
28 Sgr occultation of 1989 July 3, which sampled the equatorial region
from 6{\decdegree}6 N--15{\decdegree}2 S latitude
\citep{HubbardEtal1997icsatmes}.  There is a single profile for the
north polar region \citep[82{\decdegree}5--85{\degrees}
N]{CoorayEtal1998icnpsatoc}, and a northern low-latitude profile from
the same event \citep[19{\decdegree} 16 N]{FrenchEtal1999phemu97satoc}.
Saturn's central flash probes much deeper, around 2.5 mbar;
\citet{NicholsonEtal1995icsatcfl} obtained IR images of the flash
during the 28 Sgr event, from which they inferred the zonal wind
profile of the sampled latitudes along Saturn's limb.

Occultations observed by a spacecraft near a giant planet probe the
troposphere from the cloud deck (\sim1 bar) to the mbar level at radio
and infrared wavelengths.  They probe the upper stratosphere and
thermosphere (\math{<}1 {\micro}bar) in the ultraviolet.  Earth-based
visual and infrared occultations measure the thermal structure of the
intervening mesosphere and stratosphere regions, which are not well
sampled by spacecraft experiments.  For Saturn, the {\em Pioneer} radio
\citep{KlioreEtal1980scipioneersatoc, LindalEtal1985ajvoysatoc} and
{\em Voyager 2} extreme ultraviolet solar and stellar observations
\citep{SmithEtal1983jgrvoysatoc} sensed the equatorial region only.
The {\em Voyager 1} radio occultation sensed 75{\degrees} S
\citep{TylerEtal1981scisatvoy1radoc}, while the {\em Voyager 2} radio
occultations sensed 36{\decdegree}5 N and 31{\degrees} S
\citep{TylerEtal1982scivoy2radoc}.  The {\em Cassini} radio experiment has
performed a number of radio occultations in the equatorial region as
well as at middle and high latitudes of Saturn
\citep{NagyEtal2006jgrCassinisatradio,
  KlioreEtal2009jgrSatOc}.  The {\em Cassini} Ultraviolet Imaging Spectrograph stellar occultation probed
the upper atmosphere at 40{\degrees} S and 66{\degrees} N
\citep{Shemansky2008cosparSatcassiniuvis}, and the {\em Cassini} Composite
Infrared Spectrometer mapped Saturn's thermal atmospheric
emission, resulting in temperature maps for both hemispheres for
pressures ranging from 0.1 mbar to about 700 mbar
\citep{FlasarEtal2005sciSatcassini}.  The {\em Cassini} Visual and Infrared
Mapping Spectrometer also has the capability of observing
spectrally resolved near-infrared stellar occultations by Saturn's
atmosphere \citep{BrownEtal2004ssrVIMS}.

Temperature profiles for a variety of atmospheres from both
occultations and in situ observations show quasi-periodic
structures that are usually attributed to propagating waves.  Waves
have been reported on Venus \citep{HinsonJenkins1995icvenusradoc},
Earth (\citealp{FrittsAlexander2003revgeogwdyn}, and references
therein), Mars \citep{CreaseyEtal2006grlMarsgw,
  FrittsEtal2006jgrMarsgwaero}, Jupiter
\citep{FrenchGierasch1974jasocwav, YoungEtal1997scijupgravwav,
  RaynaudEtal2003icjupoc, RaynaudEtal2004icjupbsco,
  YoungEtal2005icjupasi}, Saturn \citep{CoorayEtal1998icnpsatoc,
  FouchetEtal2008natSaturn}, Titan \citep{SicardyEtal1999ictit28sgr},
Uranus \citep{YoungEtal2001icuoc}, Neptune
\citep{RoquesEtal1999aanepstrat3}, and Pluto
\citep{PersonEtal2008ajWavPluto, HubbardEtal2009icPlutowav,
  ToigoEtal2010icPlutoTides}.  Both the behavior of individual waves
and the form of wave power spectra can reveal properties of the
underlying atmosphere.  For example, the forcing, propagation, and
dissipation of the waves both contribute to and depend on the sources
and sinks of energy in the atmosphere, the background thermal state,
and eddy and molecular diffusion.

On 1998 November 14, Saturn and its rings occulted GSC 0622-00345, as
predicted by \citet{BoshMcDonald1992ajsatocs}.  We obtained a
light curve for atmospheric immersion, based on infrared imaging
observations at the NASA Infrared Telescope Facility (IRTF) on Mauna
Kea, HI.  The high signal-to-noise ratio (S/N) allowed us to determine
the vertical temperature profile of Saturn's stratosphere at
55{\decdegree}5 S latitude, a region not sampled by previous stellar
occultation observations (see Figure 1 and Table 1 of
\citealp{HarringtonFrench2010apjsatoc98I}, hereafter Paper I).

Paper I presents the light curve and describes the new methods used to
acquire and derive it.  This paper presents the scientific analysis of
the light curve.  Subsequent sections cover isothermal model fits,
numerical inversions to derive the thermal profile, noise tests, local
and global wavelet spectrum analysis, a gravity-wave model based on
wavelet reconstruction, exploration of the ``universal'' power
spectrum of gravity waves, discussion of the global power spectrum,
and our conclusions.  For each analysis, we present new significance
tests that determine the effects of real (non-Gaussian) noise.

\section{ISOTHERMAL FITS}

Table \ref{isotab} presents the results of isothermal model fits to
the light curve with free and fixed baselines.  The free baselines
flank the calculated values by over 1.5% of full flux, a huge
deviation given the high accuracy of the baseline determination.  The
second fit fixes the baselines at their calculated values.  For the
latter case, the derived scale height is very close to that in the
good region of the inversion presented below.

Isothermal fits can give approximate light-curve parameters that do not
depend on many assumptions and that are unique.  However, the vastly
different values for temperature, \math{T}, and scale height,
\math{H}, between the two cases and the poorly fit baselines in the
free-baselines case indicate that isothermal models do not approximate
this atmosphere well.  Both models have long, nonzero tails that stand
well above the data (see Figure 2 of Paper I).  The light curve also
contains many spikes with amplitudes that are many times the noise
level.  The spikes and the low-valued tail are features of the
observations that are not in the model but that strongly influence
where the fit falls.  Introducing a temperature gradient to the model
might improve the fit, but that does not address the numerous spikes.
Inversion is thus the proper analytic approach.

\if\submitms y
\clearpage
\fi
\atabon\begin{deluxetable}{lcc}
\tablecaption{\label{isotab} Isothermal Fit Results}
\tablewidth{0pt}
\tablehead{
\colhead{Parameter} &
\colhead{Baselines Free} &
\colhead{Baselines Fixed}}
\startdata
Half-light time (UTC)		& 11:18:47.14 {\pm} 0.18   & 11:18:46.59 {\pm} 0.17 \\
Full flux			&      1.0157 {\pm} 0.0028 & 1 (not fit) \\
Background			&     -0.0150 {\pm} 0.0013 & 0 (not fit) \\
Scale height, \math{H} (km)	&        50.7 {\pm} 1.1    &  44.7 {\pm} 0.9 \\
Temperature, \math{T} (K)	&         159 {\pm} 4      & 140   {\pm} 3
\enddata
\end{deluxetable}\ataboff
\if\submitms y
\clearpage
\fi
\placetable{isotab}

Both here and in the inversion that follows, we use the mean molecular
mass to convert \math{H} to \math{T}.  The He/H\sb{2} volume mixing
ratio for Saturn determined by \citet{ConrathGautier2000icsathe} is in
the range 0.11--0.16.  We adopt a value of 0.135 and a CH\sb{4} mixing
ratio of \math{(4.7\pm0.2)\tttt{-3}} \citep{FletcherEtal2009icCH4},
resulting in a 
mean molecular mass of 2.35 AMU.  The uncertainty in He/H\sb{2} dwarfs
the uncertainty in any other constituent.  The value of 2.135 AMU used by
\citet{HubbardEtal1997icsatmes} and many prior workers is much
smaller.  One must be careful to adjust temperatures and adiabatic
lapse rates to the same mean molecular mass when making comparisons.

\if\submitms y
\clearpage
\fi
\begin{figure*}[ht]
\if\submitms y
  \setcounter{fignum}{\value{figure}}
  \addtocounter{fignum}{1}
  \newcommand\fignam{f\arabic{fignum}.eps}
\else
  \newcommand\fignam{profiles_V13c_color.ps}
\fi
\includegraphics[height=\textwidth, clip, angle=90]{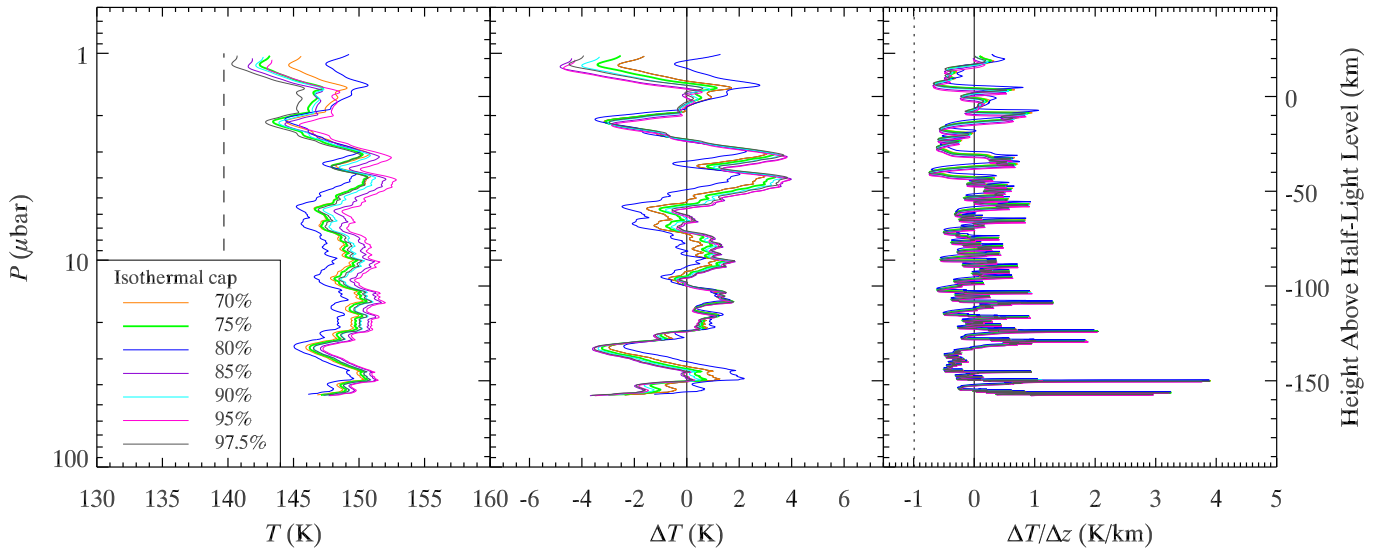}
\figcaption{\label{inversion}
Family of thermal profiles derived by numerical inversion.  The
light curve inverted for each curve has an isothermal cap that ends at
the flux level indicated in the key.  We used the 75% curve (green
line in the middle of the group) for subsequent analyses.  The data are
plotted vs.\ pressure (left axis), with the approximate corresponding
altitude scale given on the right axis.  Left: temperature profiles.
The dashed line shows the temperature of our isothermal fit.  Middle:
the same data after removal of a linear fit to each profile.  This shows
that the shape and amplitude of the small-scale temperature
fluctuations is consistent among the profiles, that there are very
small differences in their altitudes, and that the deviations from a
linear thermal profile are no more than about \pm4 K.  Right:
vertical temperature gradient vs.\ altitude, derived from the left
panel's data.  The atmosphere is statically stable over this altitude
range, since the temperature gradient is separated from the adiabatic
lapse rate (dashed line).  Nonetheless, we still see alternating
rounded and spiked structures, which are seen in many other profiles
where the vertical temperature gradient approaches adiabatic.
}
\end{figure*}
\if\submitms y
\clearpage
\fi

\section{TEMPERATURE PROFILE}

We applied an Abel transform to the normalized light curve under the
usual assumptions that the atmosphere is radially symmetric and that
ray crossing is not substantial (\citealp{FrenchEtal1978icocinv}; in
grazing occultations, unlike ours, these assumptions may be violated,
see \citealp{CoorayElliot2003apjcaustics}).  This produced \math{T},
pressure (\math{P}), and number density (\math{n}) as a function of
height above the half-light level (\math{z}).  Table \ref{invtab} and
Figure \ref{inversion} show our parameters and results.  We rebinned
the light curve prior to the inversion into time intervals
corresponding to vertical atmospheric layers of equal thickness (1 km).
This is high enough resolution to preserve the intensity spikes in the
light curve and is roughly comparable to the 1.2 km Fresnel scale
(Paper I).  Saturn is very oblate, so at any point on the surface the
radius of curvature depends on both the latitude and the direction
being considered (e.g., north-south versus east-west).  Using the
radius of curvature along the line of sight at the half-light latitude
accounts for the planet's oblateness.  The derived vertical
refractivity profile is proportional to the density profile.  The
\math{T} and \math{P} profiles then result from integrating the
hydrostatic equation and applying the ideal gas law
\citep{FrenchEtal1978icocinv}.

The inversion process determines the unique vertical temperature
profile such that a forward model of the occultation would reproduce
the observed light curve exactly, subject to assumptions just
mentioned. Unfortunately, the inversion cannot distinguish between
noise and atmosphere-induced variations in stellar intensity. Both
random and systematic noise in the observations can thus seriously
affect the derived temperature profile (see
\citealp{ElliotEtal2003ajocinv} for an extensive review, and
references cited therein).  Initially, the photometric noise in the
upper baseline completely overwhelms the actual refractive effects of
the tenuous upper atmosphere.  The noise introduces \sim100 K swings
in the upper part of the derived thermal profile, and these unphysical
swings bias the results at the onset of the inversion.  In deeper
layers, the uncertainties associated with this initial condition are
less significant.  Eventually, refractive defocussing of the starlight
is so large that the noise in the lower baseline of the light curve
dominates the faint signal.  This results in unphysical temperature
variations and trends at the deepest atmospheric layers probed during
the occultation (see, e.g., \citealp{RaynaudEtal2004icjupbsco}).  As a
rough guide, \citet{FrenchEtal1978icocinv} showed that, for
high-quality, Earth-based stellar occultations, the valid region of
the derived temperature profile ranges from about 0.5 to
-3.5\actcharon~\actcharoff\math{H} above and below the half-light
level.

Several strategies have been adopted to minimize the effects of the
unstable initial condition of the inversion process.
\citet{ElliotYoung1992ajocmodel} and \citet{ElliotEtal2003ajocinv}
explored a range of models that fit the upper part of the light curve
assuming a power-law dependence of temperature with radius.  This is
an extension of the strategy developed by
\citet{FrenchEtal1978icocinv} of replacing the upper part of the
light curve by the best isothermal fit to that restricted part of the
data.  In the absence of detailed knowledge of the upper stratospheric
temperature structure above the occultation region, we adopt the
minimalist assumption that the atmospheric region sounded by the upper
part of the light curve is isothermal.  Effectively, we assume
that the upper atmosphere above the inversion's reliable region does
not have large-scale temperature fluctuations, and is comparable in
mean temperature to the valid region.  A comparison of 28 Sgr and
{\em Voyager} UV stellar occultation observations bears this out for Saturn
\citep{HubbardEtal1997icsatmes}.  The stability of the
hybrid-light-curve inversion depends on the length of this isothermal
``cap.''  If the cap extends only from the upper baseline to the 99%
level (in units of normalized stellar flux), then the noise in the
subsequent upper part of the observed light curve will still produce
spurious temperature variations at the onset of the inversion.  As the
cap length increases, the inversion stabilizes, eventually
contaminating the inversion's valid region.  The optimal cap is large
enough to give a stable inversion but ends above the valid region.

The left panel of Figure \ref{inversion} shows the thermal structure
derived from a suite of light-curve inversions with seven different
isothermal caps ending at 97.5% -- 70% of the full stellar signal (we
computed additional caps outside this range).  The order of the
curves reflects the sensitivity of the inversion to noise and
atmospheric structure that immediately follow the isothermal cap.  The
degree of uncertainty in the inversion is shown by the spread between
the profiles.  We have used the 75% cap in all subsequent analyses.
\citet{ElliotEtal2003ajocinv} show that a 50% cap results in about a
3% error in the derived temperature at the onset of the underlying
inversion region, for an isothermal light curve with S/N per scale
height of 200 in the presence of white noise.  Since our goal is to
investigate vertical variations in temperature as well as to estimate
the mean temperature, we have adopted a compromise of replacing only
the upper 25% of the light curve by an isothermal model.

\comment{
The following table wants to shift to the right with this line:
\atabon\begin{deluxetable}{l@{\extracolsep{0.25em}}r@{\extracolsep{0.25em}}l@{\extracolsep{0.25em}}l}
It doesn't with this one:
\atabon\begin{deluxetable}{lrll}
But, we like the table produced by the first one.
}

\if\submitms y
\clearpage
\fi
\atabon\begin{deluxetable}{l@{\extracolsep{0.25em}}r@{\extracolsep{0.25em}}l@{\extracolsep{0.25em}}l}
\tablecaption{\label{invtab} Inversion Parameters and Results}
\tablewidth{0pt}
\tablehead{
\colhead{Description} &
\colhead{Value} &
\colhead{Units} &
\colhead{Comment}}
\startdata
Gravity at half-light, \math{g}	& 11.06 & m s\sp{-2} \\
Refractivity at STP\tablenotemark{a}  & 1.24\tttt{-4} \\
Mean molecular mass	  & 2.35 & AMU & \citet{ConrathGautier2000icsathe} \\
Local rad.\ of curv.\     & 64,307.9 & km & Along line of sight \\
\math{T/H}		  & 3.126 & K km\sp{-1} \\
Adiabatic lapse rate, \math{\Gamma} & -0.992 & K km\sp{-1}
\enddata
\tablenotetext{a}{standard temperature and pressure}
\end{deluxetable}\ataboff
\if\submitms y
\clearpage
\fi
\placetable{invtab}

The temperature variations are more clearly seen in the middle panel
of Figure \ref{inversion}, which shows the deviations of each of the
profiles from a linear fit to that profile.  The right panel shows the
vertical temperature gradient for each inversion.  The vertical dashed
line corresponds to the adiabatic lapse rate.  The atmosphere is
locally stable against convection from 1 -- 60 {\micro}bar.  Some
occultation temperature gradient profiles (e.g., Figure 10 of
\citealp{RaynaudEtal2003icjupoc}, Figure 7 of
\citealp{RaynaudEtal2004icjupbsco}, and others cited therein) have
oscillations with a rounded shape on the low \math{\Delta T/\Delta z}
side and a narrow, spiked shape on the high side, and so does ours,
particularly at depth.  The asymmetric rounding has been attributed to
gravity wave breaking as the profile's gradient approaches the
adiabatic lapse rate.  Saturn's adiabatic lapse rate is separated from
the negative-side extrema in our profile's gradient by 0.2 K km\sp{-1}
everywhere, and generally by much larger amounts.   One might
expect that, as waves propagate vertically and increase in amplitude
by virtue of energy conservation, they will eventually become 
superadiabatic and result in wave breaking, but this is not always
reflected in the retrieved vertical temperature profiles.  Similar
results to ours have been found for Pluto (see Figure 4 of
\citealp{YoungEtal2008ajPluto}).  The occultation
profile integrates over a large atmospheric path length, so there may
be local gradient instabilities not seen in occultation inversions.

\atabon\begin{\widedeltab}{cccccccccccccc}
\tablecaption{\label{table:noise} Inversion Set Temperature Ranges and Power-law Fits to Global Wavelet Spectra}
\tablewidth{0pt}
\tablehead{
\colhead{Sample}                     &
\colhead{Temp.}                      &
\multicolumn{2}{c}{Temp. Amplitude}  &
\multicolumn{2}{c}{Temp. Amplitude}  &
\multicolumn{2}{c}{Mean-Norm T Pow}  &
\multicolumn{2}{c}{Mean-Norm T Pow}  &
\multicolumn{2}{c}{BG-Norm Temp. Pow} &
\multicolumn{2}{c}{BG-Norm Temp. Pow} \\
\colhead{Set}            &
\colhead{Range}          &
\multicolumn{2}{c}{Mean} &
\multicolumn{2}{c}{Max}  &
\multicolumn{2}{c}{Mean} &
\multicolumn{2}{c}{Max}  &
\multicolumn{2}{c}{Mean} &
\multicolumn{2}{c}{Max} \\
                 &
\colhead{Mean}   &
\colhead{Const.\tablenotemark{a}} &
\colhead{Exp.}   &
\colhead{Const.} &
\colhead{Exp.}   &
\colhead{Const.} &
\colhead{Exp.}   &
\colhead{Const.} &
\colhead{Exp.}   &
\colhead{Const.} &
\colhead{Exp.}   &
\colhead{Const.} &
\colhead{Exp.}   \\
              &
\colhead{(K)} &
\colhead{(K)} &
              &
\colhead{(K)} &
              &
\colhead{(K)} &
              &
\colhead{(K)} &
              &
\colhead{(K)} &
              &
\colhead{(K)} &
}
\startdata
Data\tablenotemark{b}
     &  9.85 & 1.23\tttt{-1} & -1.05 &               &       & 5.78\tttt{-6} & -3.02 &               &       & 5.44\tttt{-6} & -3.05 &               &       \\
RN   &  5.52 & 2.05\tttt{-2} & -1.26 & 4.12\tttt{-2} & -1.23 & 2.76\tttt{-7} & -3.23 & 1.16\tttt{-6} & -3.00 & 2.86\tttt{-7} & -3.22 & 1.25\tttt{-6} & -2.96 \\
RS   &  1.60 & 3.71\tttt{-3} & -1.26 & 1.51\tttt{-2} & -1.22 & 2.58\tttt{-8} & -2.70 & 4.96\tttt{-7} & -2.41 & 2.69\tttt{-8} & -2.69 & 5.54\tttt{-7} & -2.40 \\
GN   &  2.72 & 1.79\tttt{-2} & -1.02 & 3.93\tttt{-2} & -0.99 & 1.52\tttt{-7} & -2.91 & 6.16\tttt{-7} & -2.83 & 1.56\tttt{-7} & -2.91 & 6.33\tttt{-7} & -2.83 \\
GS   &  0.84 & 3.27\tttt{-3} & -1.09 & 7.58\tttt{-3} & -1.07 & 7.19\tttt{-9} & -2.90 & 3.32\tttt{-8} & -2.81 & 7.40\tttt{-9} & -2.90 & 3.42\tttt{-8} & -2.81
\enddata
\tablenotetext{a}{Const.\ and Exp.\ refer to the constant and exponent (power)
of the power-law fits (Section \ref{sec:powspec}).}
\tablenotetext{b}{Range is average range for different caps.  Power law
is for 75% cap.}
\end{\widedeltab}\ataboff
\if\submitms y
\clearpage
\fi
\placetable{table:noise}

\section{NOISE TESTS}

It is not a simple matter to quantify in detail the effect of
light-curve noise on the numerical inversions.  The inferred structure
at a given atmospheric level is contaminated by errors in the derived
refractivity of all overlying levels.  \citet{FrenchEtal1978icocinv}
showed that the correlation length scale of the inversion process
extends to well over a scale height above any given pressure level, as
dictated by the width of the kernel in the integral equation for the
Abel transform.  Thus, even uncorrelated white noise in the light curve
results in correlated errors in the derived thermal profile.

The problem is even more complex if the input light-curve noise is
correlated, and almost {\em any} baseline drift on time scales longer
than the exposure time will introduce significant correlation.  Causes
of such drifts include turbulence and waves in Earth's atmosphere
(``seeing''), atmospheric transparency variations, and pointing
drifts.  The latter can be problematic in spacecraft instruments with
few spatial channels (e.g., few pixels) or with pointing-dependent
sensitivity.

Although previous investigators have explored the consequences of
white noise to the {\it mean} temperature determined by inversion
(e.g., \citealp{FrenchEtal1978icocinv, ElliotEtal2003ajocinv}), there
has been no systematic study of the effects of noise on wave analyses,
and little consideration of correlated noise.  To assess the
significance level of features in our derived profiles and subsequent
analyses, we added both Gaussian and real noise to an isothermal model
light curve.  We created 25 realizations of Gaussian noise with the
same standard deviation as our upper baseline.  To eliminate the
baseline uncertainty issue, we added this noise only below the 75%
light level.  For the real noise tests, we took a section of our upper
baseline, removed a low-order polynomial, and repeated the section
several times.  The polynomial ensures that the sections have zero
mean and meet without a discontinuity.  We shifted the resulting data
vector by three different amounts so that specific noise spikes would
appear at three different locations along the synthetic light curve.
We added this to the isothermal curve starting at each of the cap
levels to create 21 sets (seven caps times three shifts).

Sources of scintillation noise include both the star and the residual
from template subtraction (Paper I).  To account for the decrease in
the stellar contribution to scintillation, we created two additional
sample sets, identical to those above but with the noise scaled by the
normalized intensity of the noise-free isothermal model light curve.
Most ground-based occultation data sets have similar noise levels on
both baselines, indicating that the dominant noise source is residual
planetary light and that the unscaled noise analysis is most
appropriate for that case.  However, reductions in the residual
planetary signal are possible in the future, in which case the scaled
noise analysis would apply.  We label the four sets of noisy
isothermal profiles ``RN'' (real, normal), ``RS'' (real, scaled),
``GN'' (Gaussian, normal), and ``GS'' (Gaussian, scaled).

We inverted all of these light curves and calculated the temperature
ranges in the valid regions of each resulting profile.  The first
columns of Table \ref{table:noise} present the average of all the
temperature ranges in each sample set and in the observations.  Both
types of noise induced large oscillatory structures into profiles
that, without noise, should have been straight vertical lines, but
real noise had a dramatically larger effect than Gaussian noise.
Temperature ranges were typically 2 -- 3 K for the GN set, although a
single outlier had a 6.4 K range.  Within a given shift of the RN set,
the profiles and their ranges were mostly similar.  Ranges for the
three shifts averaged 7.4, 5, and 4.5 K, indicating that the placement
of individual light curve spikes strongly affected the results.  The
scaled sets had much smaller ranges, but otherwise behaved similarly
to their unscaled brethren.

The large-amplitude oscillations in these supposedly isothermal
inversions generally have wavelengths of at least a scale height.
Real noise is substantially worse than Gaussian noise of the same
standard deviation, likely because of its red power spectrum.  We
assess the power spectrum of the noise and related error and
significance issues in Section \ref{sec:powspec}.

\if\submitms y
\clearpage
\fi
\begin{figure*}[t]
\if\submitms y
  \setcounter{fignum}{\value{figure}}
  \addtocounter{fignum}{1}
  \newcommand\fignam{f\arabic{fignum}.eps}
\else
  \newcommand\fignam{satoc98-wavelet.ps}
\fi
\centerline{\includegraphics[width=0.84\textwidth, clip]{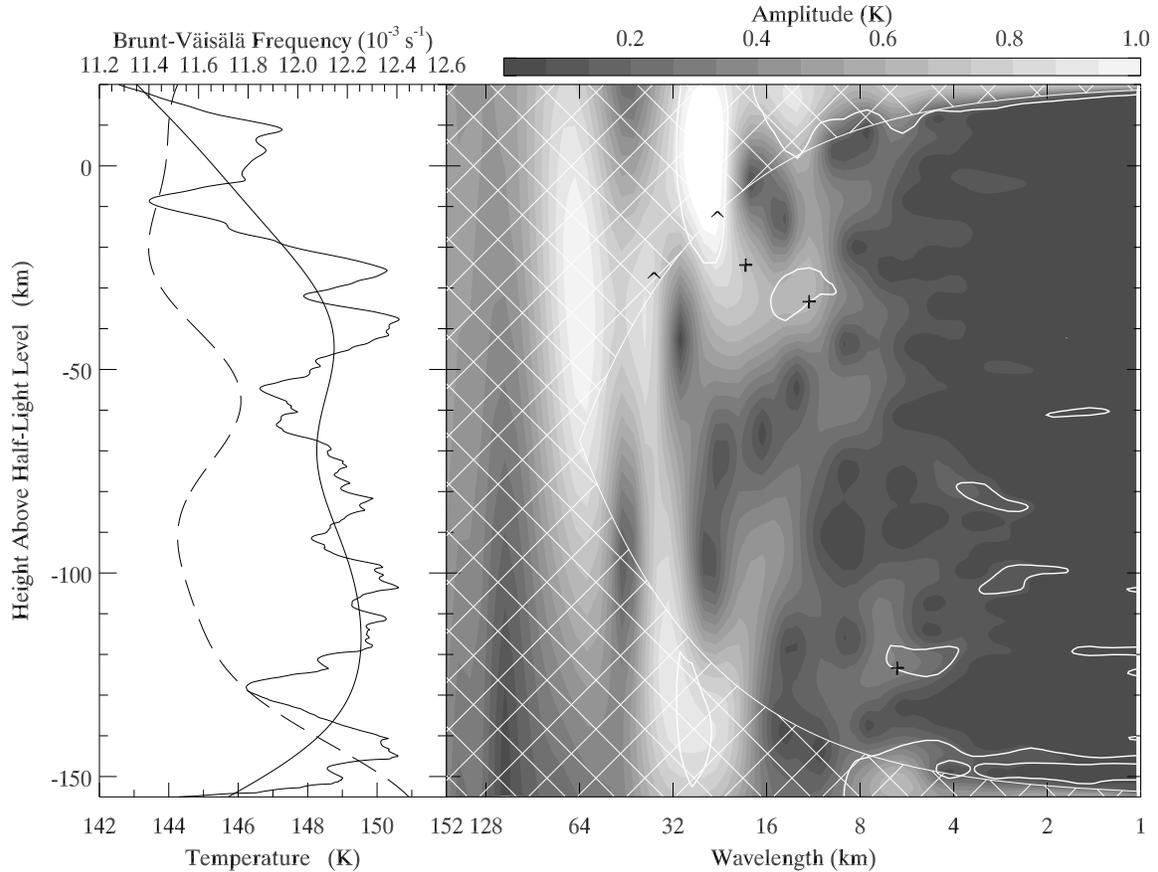}}
\figcaption{\label{waveletfig}
Wavelet amplitudes of the thermal profile.  Left: thermal profile from
the 1998 November 14 occultation by Saturn (spiky, solid line, bottom
axis), background profile used for gravity-wave modeling and the ``B''
case of the global power spectrum, below (smooth, solid line, bottom
axis), and Brunt-V{\"a}is{\"a}l{\"a} frequency derived from background
profile (dashed line, top axis).  The background profile is a wavelet
reconstruction of the thermal profile using only scales larger than 60
km.  Right: the corresponding unnormalized wavelet amplitude (not power)
spectrum, with the same vertical axis.  The horizontal axis is the
Fourier wavelength equivalent to the wavelet scale at each location.
The cross-hatched region is the unreliable ``cone of influence'' (COI)
of edge effects that alter the amplitudes.  The gray level at each
point is proportional to the temperature amplitude of a wave at that
altitude and wavelength (see scale bar at top, which omits levels in
the COI for readability).  Empirical tests with real and Gaussian
noise produced features (at larger periods) at least as strong as the
0.7-K peak at 12.5 km wavelength, -35 km height.  Symbols mark the
locations of amplitude peaks fit by gravity-wave models (Table
\ref{table:wave_modes}).  White contours encircle regions of greater
than 95% confidence per point that the signal stands significantly
above the data's global power spectrum.  See Section \ref{sec:wavelets}
for discussion, including significance contours.
}
\end{figure*}
\if\submitms y
\clearpage
\fi

\section{WAVELET ANALYSIS}
\label{sec:wavelets}

The observed temperature profiles all show fluctuations with
amplitudes as large as 4 K.  Small-scale, quasi-periodic structures in
atmospheric profiles are often interpreted as inertia-gravity waves
\citep{FrenchGierasch1974jasocwav, YoungEtal1997scijupgravwav,
  CoorayEtal1998icnpsatoc, RaynaudEtal2003icjupoc,
  RaynaudEtal2004icjupbsco}.  To investigate the wave nature of these
structures in more detail, we computed the wavelet transform of the
valid region.  A wavelet transform gives the amplitude or power
spectrum as a function of atmospheric depth.
\citet{TorrenceCompo1998bamswavelet} provide a quantitative and
accessible wavelet tutorial with software; see Acknowledgements for
Web address.  In a wavelet image of amplitude or power versus
wavelength and height, such as we present in Figure \ref{waveletfig}, a
wavetrain of constant wavelength and amplitude would appear as a
constant-brightness, vertical band.  If the wavelength varied along
the train, the ridge would tilt or curve.  Changes in amplitude would
appear as varying brightness along the band.

Following \citet{RaynaudEtal2003icjupoc, RaynaudEtal2004icjupbsco}, we
used the Morlet wavelet with a nondimensional frequency of 6
(\math{\omega\sb0} in Equation (1) of
\citealp{TorrenceCompo1998bamswavelet}).  The left panel of
Figure \ref{waveletfig} shows the valid region of the temperature
profile, for which we computed the wavelet amplitudes presented in the
right panel.  \citet[their Equation (8)]{TorrenceCompo1998bamswavelet}
provide a power spectrum ``normalization'' that allows direct
comparison to Fourier analyses, which we do below.  This adjusts for
the wavelet spectrum's geometric spacing between frequencies and
enables definite integrals of power spectral density (PSD) to compute
total power in a wavelength range.  Since amplitude spectra are not
spectral densities (i.e., one does not integrate them), one must
remove the normalization to express amplitudes in Kelvins.  We thus
keep the adjustment for the power spectra presented herein, but remove
it for amplitude spectra.  This adjustment and the orthogonal basis
set of the transform allow us to recover accurately the amplitudes of
synthetic sinusoidal signals inserted into the input data.  The
cross-hatched region is the so-called cone of influence (COI) of the
edges of the data.  In this region, points are close enough to the
edge of the data that wavelets at those periods extend beyond the
data.  This effectively averages in zeros from outside the data.
Structure within the COI is unreliable, so we ignore it.  The profile
is dominated by structures with wavelengths longer than 32 km.
Few-kilometer, irregular wiggles are superposed, but consistent strength in
scales of 5--30 km is absent.  This is reflected in the wavelet
transform, where amplitudes drop substantially outside the COI.

We also computed the global wavelet power spectrum
\citep[Equation (22)]{TorrenceCompo1998bamswavelet} by averaging the
normalized wavelet power transform over all valid heights that are
outside the COI (see Figure \ref{pscompfig}).  This method of computing
the power spectrum has advantages over both a single Fourier transform
with a window applied to the data and a similar average over a
windowed Fourier transform (WFT).  The WFT is a local transform
similar to a wavelet transform, but its sliding window has a fixed
width that affects the computation differently at each wavelength,
introducing wavelength-dependent behavior.  The wavelet transform's
window size scales with the wavelength and thus has the same effect at
each wavelength.  The local nature of both the WFT and wavelet
transform allows one to use the COI to omit points contaminated by
edge effects.  The result is a dramatic reduction in power spectrum
noise, as shown in Figure \ref{pscompfig}.  The new significance
test for the global power spectra (see Section \ref{sec:powspec}) could
not be meaningfully applied without the superior noise rejection of
wavelets.

\if\submitms y
\clearpage
\fi
\begin{figure}[t]
\if\submitms y
  \setcounter{fignum}{\value{figure}}
  \addtocounter{fignum}{1}
  \newcommand\fignam{f\arabic{fignum}.eps}
\else
  \newcommand\fignam{satoc98-pscomp.ps}
\fi
\includegraphics[width=\columnwidth, clip]{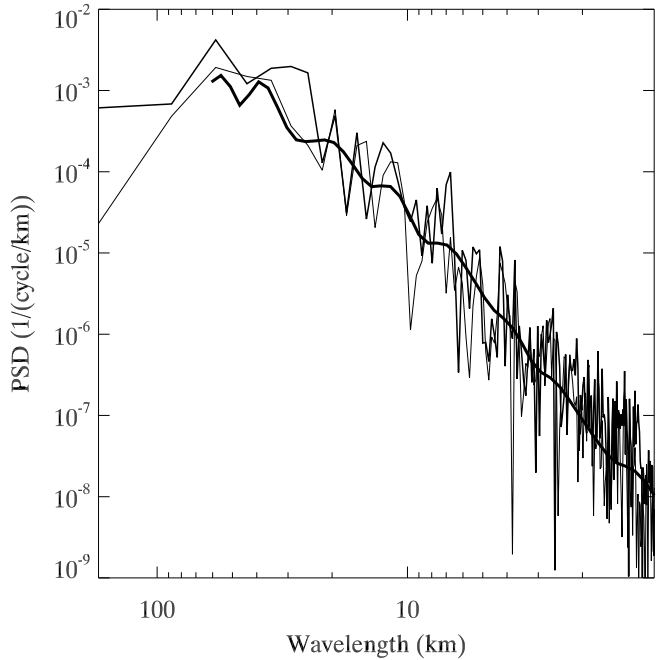}
\figcaption{\label{pscompfig}
Power spectrum of thermal profile (75% cap, B case of normalized
temperature, below) derived in three ways.  Thick line: wavelet spectrum
averaged outside of COI.  Medium line: Fourier transform with Hann
window and 8/3 scaling used, e.g., by \citet{YoungEtal2005icjupasi}.
Thin line: Fourier transform without window or scaling.  The
wavelet-derived spectrum is much smoother than either
Fourier-transform-derived spectrum.  The non-Hann Fourier spectrum
rises above the other two curves at short wavelength and falls below
the Hann version at long wavelength.  The high noise level of the
Fourier-transform-derived spectra would make them unsuitable for use
in our significance test.
}
\end{figure}
\if\submitms y
\clearpage
\fi

The data's global spectrum shows a downward-curving trend with
superposed peaks.  The general trend appears to follow an atmospheric
gravity-wave spectrum (see Section \ref{sec:powspec}).  The peaks could be
discrete gravity waves, which we study with a model (see Section
\ref{sec:gravmod}).  Finally, we compare the global spectra of the
data and
of our noise tests to establish significance (see Section \ref{sec:powspec}).

Are the features in the global and local wavelet spectra significant?
This is really two separate questions: does the global spectrum stand
significantly above the noise? and do local features stand
significantly above the global spectrum?  We discuss the latter here
and present the former in Section \ref{sec:powspec}.

\citet{TorrenceCompo1998bamswavelet} developed a rigorous significance
test based on a lag-1 autocorrelation noise model, which compares the
data to the data shifted by one point, on the assumption of a short
correlation length.  That test unfortunately does not apply to
occultation inversions, since inversions are correlated from any given
depth all the way to the top of the atmosphere.  Thus, the correlation
length varies as much as it possibly can, and is short only at the top
of the atmosphere.

We verified that the real and imaginary parts of the power transform
each had a Gaussian distribution at a given wavelength.  Then, we
followed code comments of \citeauthor{TorrenceCompo1998bamswavelet} in
applying the \math{\chi\sp{2}} probability distribution to determine
the multiplier for the global spectrum that gives the 95% confidence
level (we assume that if the power is significant at a given point, so
is the amplitude).  In Figure \ref{waveletfig}, contours encircle
regions with power greater than this level.  Note that these are not
contours of the amplitude transform, since the global power spectrum's
normalization varies with wavelength.

The strongest of the significant features that is completely outside
the COI has vertical wavelength \math{\lambda\sb{z}} = 12.5 km, a
maximum at \math{z = -35} km, and amplitude 0.7 K at that level.  This
feature can be interpreted as a short gravity-wave train, as can
several others at shorter wavelengths and much lower amplitudes.  It
is localized, meaning that it does not extend vertically over the
entire data set.  One can see, from the left panel in the plot, that
this feature lasts at least 1.5 cycles.  We show in
Figure \ref{gravmodfig} that it extends over four cycles.
\citet{YoungEtal2005icjupasi} also detected several short wave trains
in the Galileo Probe data for Jupiter.

Aside from their much-lower amplitude at a given wavelength, the
transforms of the noise data sets are qualitatively very similar to
that in Figure \ref{waveletfig}, except that the RS and GS sets have
decreasing power at lower altitude (as expected).  The large-amplitude
(many-K) oscillations in the noise sets generally have wavelengths of
at least a scale height, and are in the COI.  Wavelet amplitude maxima
outside the COI for the GN set averaged \sim0.13 K, and for the three
shifts of the RN set were 0.19, 0.30, and 0.34 K (these red-noise
features appeared at long wavelengths).  Maxima are somewhat smaller
for the scaled data sets, but this is less meaningful as their maxima
now come only from the high-altitude region; given enough samples we
would expect similar maxima.  In both real and Gaussian cases, about
half the tests contained substantial, isolated regions outside the COI
with amplitudes that stood above the surrounding features.  These
regions were broader for the real noise tests, often substantially
broader than the ones in Figure \ref{waveletfig}.  The maximum
transform amplitude outside the COI often appeared in these features.
As we note below, most or all of the data transform outside the COI
stands well above the noise.  Again, one assesses this best by
comparing global spectra.  The point here is that one cannot tell
signal from noise merely by looking at the pattern presented.

Numerous wave features in the literature have amplitudes smaller than
1 K, often in noisier data sets than this one (e.g.,
\citealp{RaynaudEtal2003icjupoc, RaynaudEtal2004icjupbsco}).  Such
cases likely also have significant, spurious fluctuations in their
transforms due to scintillation noise and other terrestrial
atmospheric effects, though it is impossible to tell for certain
without the test described above, as the degree of such noise is
rarely reported and time-correlated noise could conceivably have been
better for those observations than that for ours, even if random noise were
worse.  It thus becomes imperative to apply a wavelength-dependent
significance test, such as that presented above.  Had we simply used
the maximum of our noise transforms, we would have rejected all the
encircled regions of Figure \ref{waveletfig} as being below
2\math{\sigma} significance.  By the same token, it becomes difficult
to accept such detections without any significance test.  There are
two classes of features, discrete and global.  The significant,
discrete features have contours around them in Figure \ref{waveletfig}.
The significant, global features are the bumps in the thick line of
Figure \ref{pscompfig}, which we interpret with a gravity-wave model in
the next section and later compare to the noise spectra.

\if\submitms y
\clearpage
\fi
\begin{figure*}[t]
\if\submitms y
  \setcounter{fignum}{\value{figure}}
  \addtocounter{fignum}{1}
  \newcommand\fignam{f\arabic{fignum}.eps}
\else
  \newcommand\fignam{satoc98-gravmod.ps}
\fi
\includegraphics[width=\textwidth, clip]{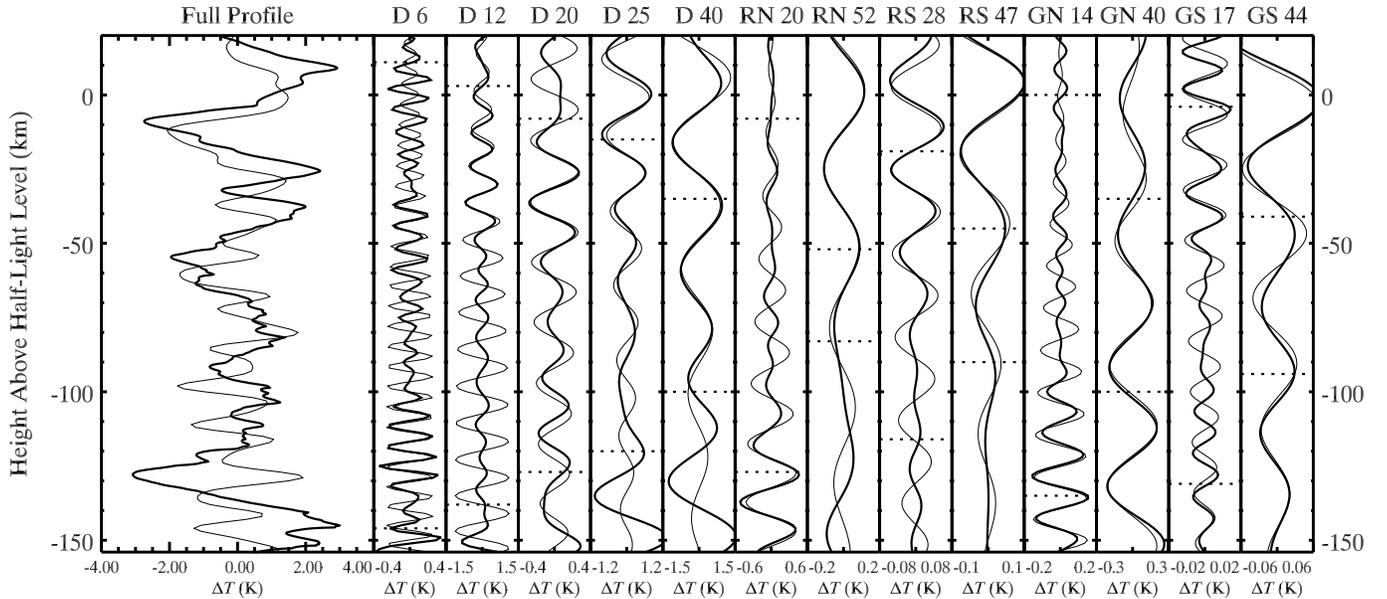}
\figcaption{\label{gravmodfig}
Gravity-wave models.  The ``Full Profile'' panel shows the 75%-cap profile
after subtracting the background state (thick line) and the sum of the
models in the next five panels (thin line).  The remaining panels
show gravity-wave models fit to these observations and to
representative profiles from the (supposedly waveless) noise sets.
The top of each panel states its source and central reconstruction
wavelength (near the wavelength at peak amplitude given in Table
\ref{table:wave_modes}), with ``D'' representing fits in the real data
shown in the leftmost panel.  In the model plots, the \comment{pair
  of} thick, solid trace\comment{s} shows the reconstructed thermal
profile around the stated wavelength.  \comment{To show the amplitude
  envelope being fit, one of these traces is 180{\degrees} out of
  phase with the data.}  The thin, solid trace gives the fitted
gravity-wave model.  \comment{The medium-weight, non-oscillating trace
  gives the amplitude envelope of this fit.}  The pair of dashed,
horizontal lines shows the altitudes where the data intersect the COI;
only information between these lines is reliable.  Parameters of the
``D'' wave fits appear in Table \ref{table:wave_modes}.  Note that,
although the RN 20 case has a relatively high amplitude for a short
stretch in the valid region, its global amplitude (mean amplitude over
all valid heights) is much lower than that of the D 20 case.  The
reconstruction for D 20 had a rather narrow range of wavelengths to
separate it from D 25, leaving significant power out of D 20.  In the
left panel, the sum of our models does not perfectly fit the data
because some models do not fit well at all altitudes and there is a
background spectrum of low-intensity activity (see Section \ref{sec:powspec}).
}
\end{figure*}
\if\submitms y
\clearpage
\fi

\atabon\begin{deluxetable}{cccccc}
\tablecaption{\label{table:wave_modes} Parameters of the Gravity Wave Fits}
\tablewidth{0pt}
\tablehead{
\multicolumn{3}{c}{Observed Parameters} & \multicolumn{3}{c}{Model-dependent Parameters}\\
\hline
\colhead{\math{\lambda\sb{z}}} &
\colhead{\math{z\sb{\rm max}}} &
\colhead{\math{\Delta T(z\sb{\rm max})}} &
\colhead{\math{z\sb{0}}} &
\colhead{\math{\lambda\sb{h}}} &
\colhead{Period} \\
\colhead{(km)} &
\colhead{(km)} &
\colhead{(K)} &
\colhead{(km)} &
\colhead{(km)} &
\colhead{(minutes)}}
\startdata
\phn6.5       & \phn   -125 & 0.26          & \phn   -114 & \phn\phn10 &  \phn13.3 \\
   12.5       & \phn\phn-35 & 0.7\phn       & \phn   -143 &    \phn120 &  \phn85.5 \\
   20\phd\phn & \phn\phn-26 & 0.26          & \phn\phn-10 &    \phn300 &     125.7 \\
   25\phd\phn &\math{>} -15 & 0.6\phn       &\math{>} -15 &    \phn100 &  \phn35.6 \\
   40\phd\phn &\math{>} -30 & 1\phd\phn\phn &\math{>} -30 &    \phn200 & \phn42.0
\enddata
\end{deluxetable}\ataboff
\if\submitms y
\clearpage
\fi
\placetable{table:wave_modes}

\section{GRAVITY-WAVE MODEL}
\label{sec:gravmod}

The noise transforms do not show the continuous vertical ridges of
elevated amplitude one would expect of a wave that propagates undamped
through the entire valid region of the temperature profile.  However,
it may be possible, by fitting an appropriate model, to improve
sensitivity to such individual gravity-wave modes in the data, using
the vertical coherence that the noise does not exhibit.
\citet{RaynaudEtal2003icjupoc, RaynaudEtal2004icjupbsco} fit such
models to identify gravity-wave signatures in Jovian temperature
profiles derived from two stellar occultations.  Note that these two
papers use different methods.  \citet{RaynaudEtal2003icjupoc} used
three chords from one event to determine both \math{\lambda\sb{z}} and
the horizontal wavelength, \math{\lambda\sb{h}}.  Through their Equation (15)
they determine the dissipation level for the wave and point out that
the observed \math{T} peaks are too high in the atmosphere to be
caused by a gravity wave with the derived parameters.  In contrast,
the event reported by \citet{RaynaudEtal2004icjupbsco} has just a
single observation.  The horizontal wavelength is a free parameter and
fits are consistent with a gravity-wave interpretation.

We follow \citet{RaynaudEtal2004icjupbsco} very closely to test
whether waves may be propagating at the wavelengths corresponding to
peaks in the global wavelet spectrum shown in Figure \ref{pscompfig}.
These correspond to regions outside the COI and at \math{\lambda\sb{z}
  < 60} km in Figure \ref{waveletfig}.  The first three columns of
Table \ref{table:wave_modes} give \math{\lambda\sb{z}}, altitude
(\math{z\sb{\rm max}}), and temperature amplitude (\math{\Delta
  T[z\sb{\rm max}]}) of five temperature peaks; these are marked in
Figure \ref{waveletfig}.  The remaining columns give the altitude of
maximum amplitude of the model wave (\math{z\sb{0}}),
\math{\lambda\sb{h}}, and the wave period.  Physically, \math{z\sb{0}}
represents the altitude where wave damping exceeds the natural
exponential amplitude growth.  The intensity of dissipative processes,
\math{\lambda\sb{z}}, and \math{\lambda\sb{h}} determine this
altitude.

As a wave propagates, variations in the atmosphere's steady-state
properties affect its amplitude and vertical wavelength.  In a
conservative atmosphere, a vertically propagating wave experiences
exponential growth of its temperature amplitude as a consequence of
the exponential decrease of \math{n} and the requirement for energy
conservation.  If the amplitude becomes large enough for the local
temperature gradient to exceed the adiabatic lapse rate,
\math{\Gamma}, the wave becomes unstable and overturns.  Dissipative
processes such as molecular viscosity, eddy diffusion, thermal
conduction, and radiative damping can further limit amplitude growth,
causing the wave to deposit its energy in the background atmosphere
\citep{Lindzen1981jgrturbstress}.  Furthermore, variations in the
background temperature and vertical shear in the zonal wind can also
cause observable changes in the amplitude, vertical wavelength, and
phase propagation.

We simulate these effects with a hydrostatic WKB gravity wave model in
a rotating atmosphere.  The model \citep{MatchevaStrobel1999icjupgw}
includes dissipation by molecular viscosity, eddy diffusion, and
thermal conduction.  We assume that there is no vertical gradient in
the zonal wind.  Calculations that included constant wind shear did
not improve the fits.  We assume scales larger than 60 km to represent
the atmosphere's steady state, although from a single occultation we
cannot determine whether an observed structure is transient or static.
Our steady-state profile is the inverse wavelet transform of scales
larger than 60 km in the right panel of Figure \ref{waveletfig}; it
appears as the smooth, solid line in the left panel.

The eddy diffusion coefficient, an input in our wave model,
parameterizes the intensity of the vertical mixing in the atmosphere.
Published values for Saturn's eddy diffusion coefficient differ by
more than 2 orders of magnitude
(\citealp{MosesEtal2000icsatphotchem} and references therein).  In our
model it varies with altitude and is proportional to \math{n\sp{-1/2}}
\citep{Atreya1986bkaips}.  We set it to 500 m\math{\sp{-2}}s at the
0.1-{\microbar} level to agree with {\em Voyager's} UVS observations
\citep{SmithEtal1983jgrvoysatoc}.  The vertical mixing in the
atmosphere is important as a wave dissipation mechanism.  The degree
of wave dissipation is determined by the eddy diffusion coefficient
and the wave's vertical and horizontal wavelengths.  Since the
temperature profile only has vertical information, the horizontal
structure is a free parameter.  The exact eddy diffusion coefficient
value is thus not critical for the wave model because adjusting the
horizontal wavelength, \math{\lambda\sb{h}}, within a reasonable range
can compensate for reasonable changes in the eddy diffusion coefficient
without changing the vertical structure.

The wave model parameters are \math{\lambda\sb{h}},
\math{\lambda\sb{z}}, temperature amplitude \math{\Delta T}, and phase
\math{\phi} at the lower boundary.  We adjust the wave parameters so
that the model fits the observations at \math{z\sb{\rm max}} (see
Table \ref{table:wave_modes}).  Since the background atmosphere's
temperature varies with altitude, so does \math{\lambda\sb{z}} for
each wave.  We select a short range in \math{\lambda\sb{z}} for each
candidate wave, zero all other scales in the wavelet transform, and
reconstruct the temperature fluctuations due just to those scales
using the inverse transform.  We compare the simulated wave and the
reconstructed temperature variations within the limits set by the COI.
The ``D'' panels of Figure \ref{gravmodfig} present the results.

For the purpose of wave identification, one would ideally like to
follow the wave signature over several wavelengths and several scale
heights, with at least two scale heights below the altitude of wave
damping.  However, the region outside the COI becomes progressively
shorter for larger wavelengths.  Therefore, large-\math{\lambda\sb{z}}
temperature fluctuations are difficult to interpret uniquely as
signatures of propagating waves.

We obtain a relatively good fit for the longest wavelength that we
consider (\math{\lambda\sb{z} = 40} km), detecting significant power
throughout the observed region.  The simulated wave has a
\math{\lambda\sb{h} = 200} km; the corresponding gravity-wave
period is 42 minutes.  The wave exists below the expected altitude of
dissipation (altitude of maximum amplitude), and shows modest
amplitude growth.  Both the amplitude and the phase of the simulated
wave follow the reconstructed temperature fluctuations well.  However,
despite the apparently good fit, we cannot rule out alternative
interpretations, since the reliable part of the reconstruction
contains only two wave cycles.

For the rest of the candidates, we have mixed success in fitting the
observed temperature fluctuations with a single propagating gravity
wave.  In general, we are able to fit the reconstructed temperature
fluctuations (both amplitude and phase) well over 2 -- 5 cycles near
the peak amplitude, but we are not successful in matching the
observations throughout the entire altitude region.
\citet{YoungEtal2005icjupasi} also present wave activity in the
Galileo Probe data for Jupiter that spans only a few cycles before it
disappears.  As previously \citep{RaynaudEtal2004icjupbsco}, our
requirement for a positive wave identification is much more
restrictive, since we require a good fit throughout the profile's
valid region.

Our wave model fit to the 25 km reconstructed scales does a relatively
good job following the amplitude of the temperature fluctuations
throughout the extent of the data, but it fails to fit the phase of
the observations at the bottom of the occultation (\math{z < -70} km).

In the case of the 20 km structure, we have a good phase match
throughout the sampled region, but the amplitude exhibits a double
peak that is difficult to explain with a simple gravity-wave model,
with or without a constant vertical wind shear.  The amplitude fits
well within four wavelengths of the top and bottom of the valid
region, but overestimates the observed fluctuations in the central
part of the profile.  A variable wind shear might explain the observed
amplitude.

The 12.5 km structure has a well defined power maximum that is nicely
separated from the COI.  This and the relatively small vertical scale
make the wave analysis more robust.  We achieve a good amplitude and
phase fit for four wavelengths for \math{z > -50} km.  At \math{z =
-50} km, the amplitude shows a very fast increase with altitude, which
is inconsistent with our model.  Such a fast amplitude variation can
be an indication of significant vertical shear in the zonal wind.  If
this were the case, however, we should also detect a sudden change in
\math{\lambda\sb{z}}, but the observed change is not very large.

The temperature reconstruction of the 6.5 km candidate shows a
significant variation in amplitude.  These variations can be modeled
with some success if one assumes that the background temperature
profile retains scales, \math{L\sb{z}}, below 60 km (e.g.,
\math{L\sb{z} > 30} km).  For our background temperature profile
(\math{L\sb{z} > 60} km), the model fits well in both amplitude and
phase for 4.5 wave cycles at altitudes below -100 km.  However, the
model does not reproduce the multiple amplitude peaks.  The
significant amplitude variation suggests beating wave modes whose
wavelengths are unresolved by the data.  The observed amplitude can be
well modeled by superposing two sine waves that have vertical
wavelengths 5.96 km and 6.66 km respectively.

For a given eddy diffusion coefficient, \math{K}, waves with short
\math{\lambda\sb{z}} and long \math{\lambda\sb{h}} dissipate low in
the atmosphere.  This places a natural filtering mechanism for
wavelengths that can propagate at a given altitude.  In order for a
wave with a short \math{\lambda\sb{z}} (e.g., 6 km) to propagate at
the probed pressure levels, \math{\lambda\sb{h}} must be rather small
(\sim 10 km).  The wave then comes close to violating the hydrostatic
approximation (\math{\lambda\sb{z} << \lambda\sb{h}}).  It also raises
a question about the detectability of such a wave by a stellar
occultation, which averages the properties of the atmosphere along the
line of sight.  The length of this averaging on Saturn is about
\math{\sqrt{ 2\pi rH} = 4500} km, where \math{r} is the planetary
radius.  \citet{SicardyEtal1999ictit28sgr} demonstrate that the
amplitude of light-curve fluctuations resulting from the presence of a
monochromatic wave with a projection of the horizontal wavelength
along the line of sight \math{l = \lambda\sb{h}/\cos \theta\sb{l}} is
not significantly reduced if
\begin{equation}
\label{eq:detectability}
{l\over \lambda\sb{z}} > \left(\frac{r}{4H}\right)\sp{0.5} = \eta,
\end{equation}
where \math{0\degrees < \theta\sb{l} < 90\degrees} is the angle
between the line of sight and the horizontal wave propagation
direction.  For Saturn, \math{\eta = 18}.  If the 6.5 km feature is to
be interpreted as a gravity wave with a horizontal wavelength
\math{\lambda\sb{h} = 10} km, the wave must be propagating at an angle
\math{\theta\sb{l} > 85\degrees} in order to be detected during the
occultation.  In other words, we would have detected a wave that
propagated almost along the planetary meridian.  In this respect, of
all the waves that might be present in the atmosphere, we
preferentially detect waves that propagate in the horizontal at large
angles \math{\theta\sb{l}}.  In summary, the arguments for a wave
interpretation of the very short scales (5 -- 8 km) present in the
temperature profile are easiest to accept if the wave field is
isotropic.

To see how well the model-fitting method rejects noise, we performed
the same analysis on one randomly selected profile from each noise
set; results are in the final eight panels of Figure \ref{gravmodfig}.
For each type of noise realization we reconstruct two scales, one
short (less than 30 km) and one long (about 40 -- 60 km), and fit
gravity-wave models.  The fits to reconstructions of noise and data
show similar qualities.  Short scales fit well over a few wavelengths
but not through the entire altitude range, and large scales fit
relatively well throughout the vertical range, but the region outside
the COI contains less than two full wave cycles.  This makes it
difficult to identify an atmospheric wave based only on the goodness
of fit.  However, there are ways to distinguish some waves from noise.
First, our detected waves have much higher amplitude than the noise at
the same scale.  The relevant scales appear significant in the global
wavelet power spectrum (Figure \ref{pscompfig}), even though most span
only a short vertical range.  Second, in the noise fits most of the
power (including the peak) is typically within the COI, and the
amplitude varies with altitude faster than expected for a wave.
Unfortunately, the latter are tendencies rather than robust
discriminators; real waves can do the same, making it difficult to
specify robust criteria to discriminate between waves and
noise-induced features.  For example the fits of the 40-km data
structure (D40) and the 40-km noise structure (GN40) look very
similar.  The amplitude of the fit to the real data is, however, more
than five times larger than the fit to the noise.  This
signal-to-noise assessment is thus the only reliable discriminator
known to us.

\if\submitms y
\clearpage
\fi
\begin{figure}[t]
\if\submitms y
  \setcounter{fignum}{\value{figure}}
  \addtocounter{fignum}{1}
  \newcommand\fignam{f\arabic{fignum}.eps}
\else
  \newcommand\fignam{satoc98-normtemp.ps}
\fi
\includegraphics[width=\columnwidth, clip]{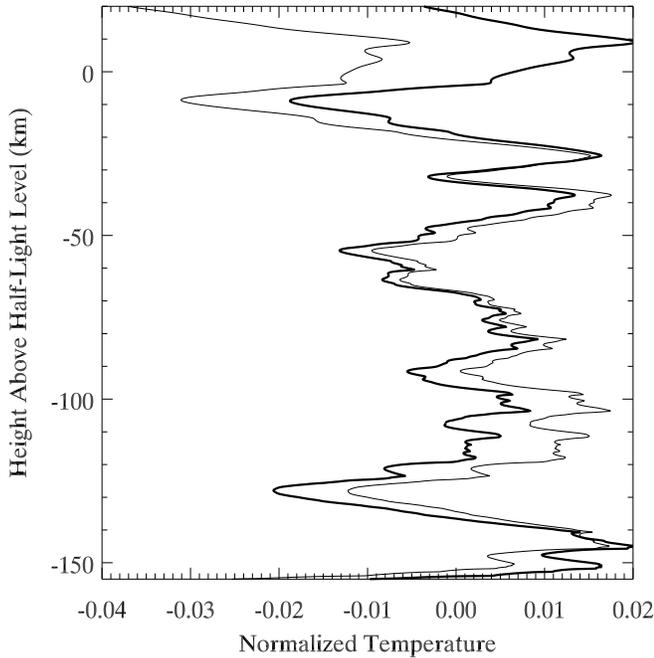}
\figcaption{\label{normtempfig}
Normalized temperature profiles derived from the 75%-cap inversion.
The thick profile uses the wavelet-smoothed background profile of
Figure \ref{waveletfig} (B case).  The thin profile uses the mean of
the temperature profile in the altitude range presented (M case).
}
\end{figure}
\if\submitms y
\clearpage
\fi

We conclude that, while this method may do well at quantifying the
properties of strong waves, it is not a good discriminator between
weak waves and noise.  A number of criteria should be satisfied for
the entire region outside the COI to make a positive wave
identification:

\begin{enumerate}
\item the wave amplitude should be many times the mean amplitude of
the worst-case real noise at that \math{\lambda\sb{z}};

\item a wave model should fit both the amplitude and phase of the
observed temperature fluctuations well;

\item the structure should have more than a few cycles.
\end{enumerate}

Note that structures failing one or more criteria may still be (or
contain) waves.  We discuss the broad spectrum of weaker gravity waves
in the next section.

\section{POWER SPECTRUM}
\label{sec:powspec}

To compare the power spectrum of temperature fluctuations in Saturn's
atmosphere with that in other atmospheres, we calculate global spectra
for normalized thermal profiles, \math{(T-\bar{T})/\bar{T}}, where
\math{\bar{T}} is the mean temperature.  While others
\citep[e.g.,][]{YoungEtal2005icjupasi, AllenVincent1995jgrgravwav}
have simply chosen regions of their data where the background was
clearly isothermal to calculate \math{\bar{T}}, we do not have this
luxury, so we computed it in two ways.  In the first, \math{\bar{T}} is
the mean temperature (148.04 K) in the good region of the profile.  In
the second, \math{\bar{T}} is the wavelet-reconstructed thermal
profile for \math{\lambda\sb{z} > 60 \mbox{ km}} (the smooth, solid
line in the left panel of Figure \ref{waveletfig}).  The choice of 60
km is arbitrary, hence the two cases.  We call them M and B, for mean
and background, respectively, and present them in
Figure \ref{normtempfig}.

\if\submitms y
\clearpage
\fi
\begin{figure}[t]
\if\submitms y
  \setcounter{fignum}{\value{figure}}
  \addtocounter{fignum}{1}
  \newcommand\fignam{f\arabic{fignum}.eps}
\else
  \newcommand\fignam{satoc98-noisespec-m.ps}
\fi
\includegraphics[width=\columnwidth, clip]{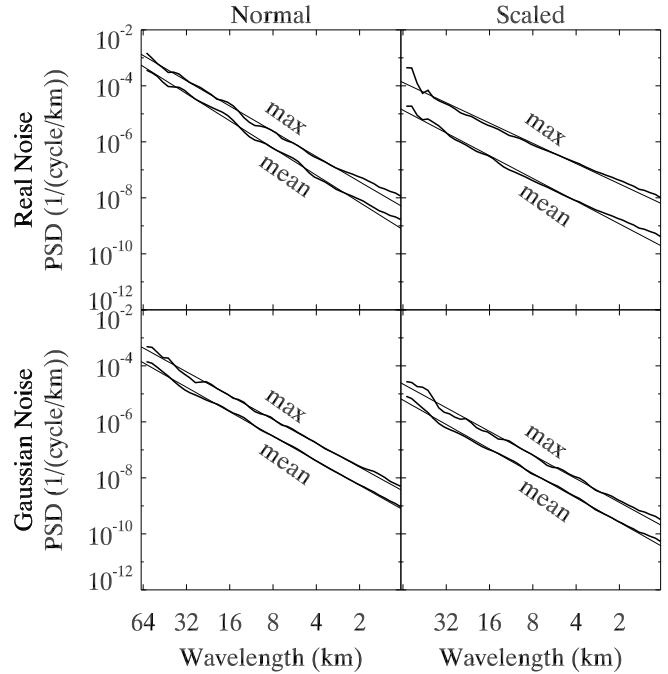}
\figcaption{\label{noisespecfig}
Power spectra of inverted noise-test light curves (thick lines), and
power-law fits (thin lines).  The words ``max'' and ``mean'' indicate
global spectra generated by those two methods from each of the four
types of noise tested (see the text).  Parameters of these fits appear in
Table \ref{table:noise}.  We compare them to the data in Figure
\ref{dataspecfig}.
}
\end{figure}
\if\submitms y
\clearpage
\fi

\if\submitms y
\clearpage
\fi
\begin{figure*}[t]
\if\submitms y
  \setcounter{fignum}{\value{figure}}
  \addtocounter{fignum}{1}
  \newcommand\fignaml{f\arabic{fignum}l.eps}
  \newcommand\fignamr{f\arabic{fignum}r.eps}
  \newcommand\figwid{0.48\textwidth}
\else
  \newcommand\fignaml{satoc98-dataspec-m.ps}
  \newcommand\fignamr{satoc98-dataspec-b.ps}
  \newcommand\figwid{\columnwidth}
\fi
\includegraphics[width=\figwid, clip]{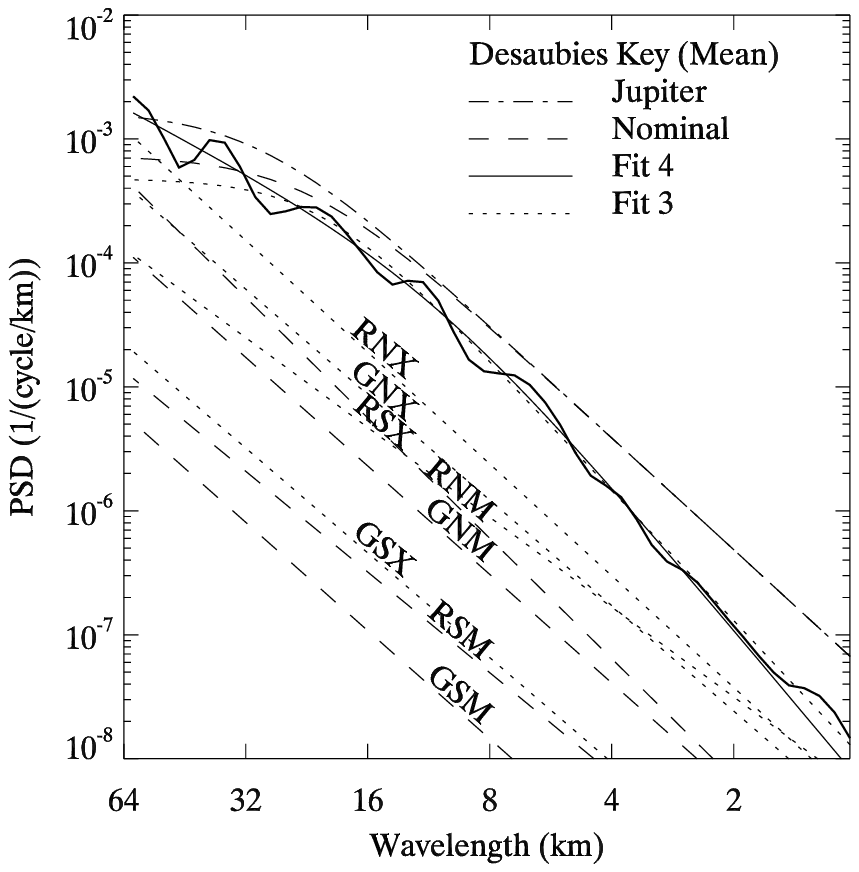}\hfill
\includegraphics[width=\figwid, clip]{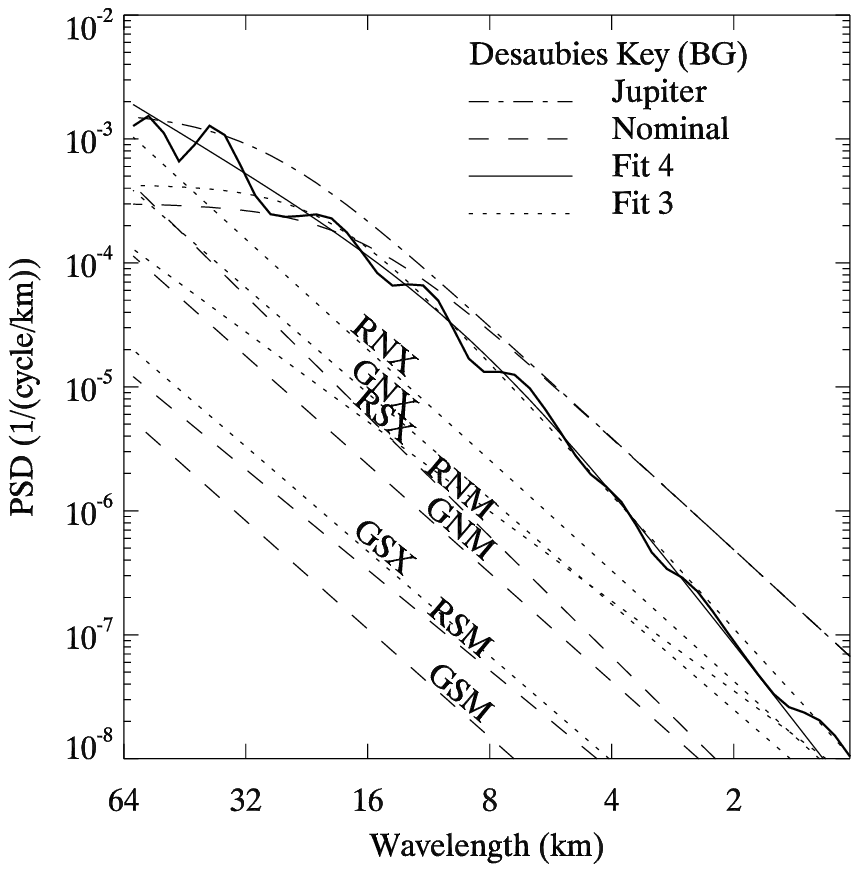}
\figcaption{\label{dataspecfig}
Global power spectrum of the data (thick line) compared to data
fits (smoothly curving lines, see key) and noise spectrum fits
(straight lines).  This compact presentation relates the main points
of this paper; see Section \ref{sec:powspec} for interpretation.
Mean-normalized data and fits are on the left, background-normalized
on the right.  The noise spectra fits are error estimates for the data
spectrum under our eight noise models for power spectra of inversions.
Table \ref{table:noise} gives parameters of these fits.  The three
characters of each noise label indicate, in order, the type of noise
set (Real or Gaussian), whether it is a Normal or Scaled set, and
whether the spectrum came from a vertical average over the maXima or
Means of the power at each location in the wavelet spectra of the
inversions in that noise set.  For example, ``RNX'' means Real noise,
Normal (non-scaled), averaged vertically over the maXima.  Mean
spectra have dashed lines; maximum spectra have dotted lines.
Standing many times higher than the linear noise
spectra, the data show peaks superposed on the modified Desaubies
function (Equation (\ref{eq:desaubies})) that describes the ``universal
spectrum'' of gravity waves for Earth.  The four smoothly curving
traces are the modified Desaubies fits given in the key and Table
\ref{table:desaubies}. The peaks are the discrete waves shown in
Figure \ref{gravmodfig}, analyzed in Section \ref{sec:gravmod}, and
presented in Table \ref{table:wave_modes}.  They cause the large
reduced-\math{\chi\sp{2}} values of Table \ref{table:desaubies}.
}
\end{figure*}
\if\submitms y
\clearpage
\fi

From their wavelet transforms, we derived the global power spectrum
for both normalized temperatures from our nominal profile (75% cap),
as described in Section \ref{sec:wavelets}.  To derive noise spectra,
we calculated the wavelet spectrum of each noise inversion, then found
the mean power at each wavelength over all the altitudes and
inversions in each set, excluding points in the COI.  Since we
demonstrated above (Figure \ref{gravmodfig}) that real noise can induce
wave-like features, we also derived spectra based on the maximum
rather than the mean, to ensure that anything we call real must stand
well above these spurious features.  Figures \ref{noisespecfig} and
\ref{dataspecfig} and Table \ref{table:noise} present the resulting
power spectral densities and power-law fits, computed with the full,
wavelet-normalized power spectrum for comparison to other work.  The
table also gives power-law fits to temperature amplitude, with neither
wavelet nor temperature normalizations applied.

The noise sets' power spectra follow an \math{m\sp{-3}} dependence,
where \math{m} is the wavenumber, that is very consistent along their
entire lengths, save for small tails near the Nyquist frequency.  The
tails are well below the resolution of the uninterpolated light curve,
so we exclude them from the fits.  The non-white noise should be due
almost entirely to Earth's atmosphere.  The \math{m\sp{-3}} power
law is unfortunate, as several atmospheric processes share it.  These
include the short-scale end of the universal gravity-wave spectrum
observed on Earth and Jupiter (see below) and short-scale,
two-dimensional (2D) turbulence \citep{HarringtonEtal1996aicjtteii,
  Travis1978jasVenusatm}, although 2D dynamics, at least, likely break
down at scales larger than are relevant to the current observations.
The scaled noise sets had lower power (see Figure \ref{noisespecfig}
and constants in Table \ref{table:noise}), but otherwise behaved
similarly to their unscaled brethren.  The spectra computed by taking
the maximum lay a factor of 4 or more above those computed by
taking the mean, which the fit constants of Table \ref{table:noise}
reflect.

Figure \ref{dataspecfig} presents the data and fits to the noise
spectra (and spectrum models; see below).  At all wavelengths, the
power spectrum of the data is 2 -- 10 times higher than even the most
conservative noise spectrum (labeled ``RNX'') and departs from the
very linear shape of the noise spectra.  We can confidently accept
that the data's spectrum reflects properties of Saturn's atmosphere
rather than noise.  The log plot allows by-eye S/N calculation under
different noise models.  For example, at a wavelength of 32 km, the
data PSD is \ttt{-3} and the RNX noise spectrum is 2\tttt{-4},
indicating a 5\math{\sigma} detection of the planetary spectrum there.
Under GNX noise it would be 10\math{\sigma}, and under GNM noise it
would be a 50\math{\sigma} detection.  We thus conclude that
comparison to real, not Gaussian, noise is necessary and that
averaging over the maximum power at each wavelength and location in
the noise set (rather than the mean) is necessary because of the
spurious, wave-like features induced by real noise.

We can now investigate the spectrum of gravity waves on Saturn, though
with the caveat that temperature profiles derived from occultations
give an average over a long horizontal path through the atmosphere
(see above).  On Earth, the spectrum of temperature fluctuations is
constant enough over season and location to be called the ``universal
spectrum'' \citep{VanZandt1982grlUnivspec,
  BalsleyCarter1982grlatmfluc, DewanEtal1984grlVelspec,
  Vincent1984jatpGwtherm, SmithEtal1987jasSatspecgw,
  TsudaEtal1989jasSatgwspec}.  It has been observed on Jupiter
\citep{YoungEtal2005icjupasi} as well, and follows the modified
Desaubies function,
\begin{equation}
\label{eq:desaubies}
p(m) = a \frac{N\sp{4}}{g\sp{2}m\sp{3}\sb{*}}
         \frac{(m/m\sb{*})\sp{s}}{1+(m/m\sb{*})\sp{s+t}},
\end{equation}
where \math{p} is the PSD,
\math{a} is a unitless constant,
\math{N} is the Brunt-V{\"a}is{\"a}l{\"a} frequency,
\math{m\sb{*} = |\Gamma|/2\sigma\sb{T} = 2\pi/L\sb{*}} is the critical
wavenumber,
\math{L\sb{*}} is the critical wavelength,
\math{\sigma\sb{T}} is the rms temperature fluctuation
of \math{(T-\bar{T})},
\math{s} is the long-wavelength power-law exponent, and
\math{t} is the short-wavelength exponent.  

\atabon\begin{\widedeltab}{lcccccc}
\tablecaption{\label{table:desaubies} Parameters of the Modified
  Desaubies Fits}
\tablewidth{0pt}
\tablehead{
\colhead{Case} &
\colhead{\math{a}} &
\colhead{\math{m\sb{*}}} &
\colhead{\math{s}} &
\colhead{\math{t}} &
\colhead{\math{L\sb{*}}\tablenotemark{a}} &
\colhead{\math{\chi\sp{2}}/dof\tablenotemark{b}} \\
\colhead{} &
\colhead{} &
\colhead{(km\sp{-1})} &
\colhead{} &
\colhead{} &
\colhead{(km)} &
\colhead{}
}
\startdata
M Jupiter       & 0.1000 {\phfz}      & 0.207 {\phrz}     & \phn0.00 {\phtz}    & 3.00 {\phtz}    &    30.3 {\phoz}   &    780 \\
B Jupiter       & 0.1000 {\phfz}      & 0.207 {\phrz}     & \phn0.00 {\phtz}    & 3.00 {\phtz}    &    30.3 {\phoz}   &    791 \\
M nominal       & 0.1000 {\phfz}      & 0.273 {\phrz}     & \phn0.00 {\phtz}    & 3.00 {\phtz}    &    23.0 {\phoz}   &    720 \\
B{\phd} nominal & 0.1000 {\phfz}      & 0.367 {\phrz}     & \phn0.00 {\phtz}    & 3.00 {\phtz}    &    17.1 {\phoz}   &    681 \\
M       fit 4   & 0.1098 {\pm} 0.0016 & 0.564 {\pm} 0.043 &    -1.71 {\pm} 0.09 & 3.92 {\pm} 0.06 &    11.1 {\pm} 0.9 & \phn17 \\
B{\phd} fit 4   & 0.1089 {\pm} 0.0015 & 0.725 {\pm} 0.056 &    -1.95 {\pm} 0.07 & 4.23 {\pm} 0.08 & \phn8.7 {\pm} 0.7 & \phn13 \\
M       fit 3   & 0.0848 {\pm} 0.0015 & 0.299 {\pm} 0.006 & \phn0.00 {\phtz}    & 3.49 {\pm} 0.02 &    21.0 {\pm} 0.4 & \phn21 \\
B{\phd} fit 3   & 0.0872 {\pm} 0.0016 & 0.313 {\pm} 0.006 & \phn0.00 {\phtz}    & 3.58 {\pm} 0.02 &    20.1 {\pm} 0.4 & \phn19
\enddata
\tablenotetext{a}{\math{L\sb{*} = 2 \pi / m\sb{*}}.}
\tablenotetext{b}{dof = degrees of freedom.  Note that we fit only the
  broadband spectrum, not the waves.}
\end{\widedeltab}\ataboff
\if\submitms y
\clearpage
\fi
\placetable{table:desaubies}

The characteristic wavenumber \math{m\sb{*}} divides the spectrum into
two regions: a small wavenumber region (\math{m<m\sb{*}}) and a large
wavenumber region (\math{m>m\sb{*}}).  The shape and magnitude in the
small-wavenumber region (large vertical wavelengths) is believed to be
dominated by the wave source characteristics and is proportional to
\math{m\sp{s}}.  The exponent \math{s} for the terrestrial atmosphere
is not well constrained, though it is typically about 1.  In
the large-wavenumber regime (small wavelengths) saturation and/or
dissipation processes are believed to control the wave spectrum
(\citealp{Gardner1996jatpGwsatdoss} and references therein).  In this
region the spectrum is proportional to \math{m\sp{-t}}, where
\math{t=3}.  The overall amplitude of the spectrum is controlled by
the dimensionless parameter \math{a}, which may represent the wave
generation mechanism.

\citet{YoungEtal2005icjupasi} applied terrestrial parameters \math{a =
  0.1}, \math{s = 0}, and \math{t = 3}, and derived \math{N = 0.0176}
s\sp{-1}, \math{\Gamma = 2.11} K km\sp{-1}, and \math{\sigma\sb{T} =
  5.0} K to compute \math{L\sb{*} = 30.3} km and thus the Desaubies
function for Jupiter.  Note that in their paper, a typographical error
gives \math{m\sb{*}=\Gamma\sigma\sb{T}/2}, but the value is calculated
correctly.  Their Equation (3) gives a curve that lies a factor of
\math{\sim2\pi} lower than the curve presented in their Figure 7, using
the parameters given above, though the shape is the same.  It is
unclear whether the plotted data have been similarly shifted or
whether the two traces do not in fact follow one
another. \comment{Asked Leslie in 2005, but she didn't reply after
  several reminders and promises she would.}

Figure \ref{dataspecfig} presents Equation (\ref{eq:desaubies}) evaluated for
the parameters given in Table \ref{table:desaubies}.  Some of these
curves have free parameters fit with a Levenberg-Marquardt minimizer;
the uncertainties on those parameters are from the minimizer's
covariance matrix.  The ``Jupiter'' case in Table
\ref{table:desaubies} and Figure \ref{dataspecfig} uses \math{m\sb{*}}
from \citet{YoungEtal2005icjupasi}, but \math{N = 0.01165} s\sp{-1}
and \math{\Gamma = 0.992} K km\sp{-1}, appropriate to our data.  The
poor visual fit and high \math{\chi\sp{2}} show that it misses the
data by a substantial margin.  Using \math{\sigma\sb{T}} calculated
from the B and M cases (1.18 and 1.98 K, respectively) yields the
\math{m\sb{*}} and \math{L\sb{*}} given in the ``nominal'' cases
(i.e., Saturnian values and no free parameters).  Things are better,
but not good.  However, the long atmospheric path of the occultation
ray likely reduced \math{\sigma\sb{T}} well below what would be
observed in-situ; for the Galileo Probe data, \math{\sigma\sb{T}
  = 5.0} K, for example.  For the case where all parameters vary,
\math{a} is still comfortably close to 0.1.  Fixing \math{s} = 0 puts
\math{a} closer to 0.085.  In all cases with free exponents, the
exponents move substantially away from their nominal values.  In
particular, \math{t} is now decidedly steeper than the -3 power law
usually cited for gravity wave spectra.  Since waves with small
\math{\lambda\sb{z}} are more likely to have small
\math{\lambda\sb{h}}, the steeper power law may be due to preferential
filtering of short wavelengths due to averaging along the occultation
ray path.

The reduced \math{\chi\sp{2}} is still high, indicating that the
undulations in the spectrum are significant: there is more here than a
background spectrum of gravity waves.  The gravity-wave model
presented above fits averages over the sections of the wavelet
transform that produce these bumps.  That model assumes a single
wavelength per wave, so we cannot use its output to make a complete,
simultaneous model of background and discrete waves that might fit the
global spectrum more perfectly.

\section{DISCUSSION OF THE SPECTRUM}

The one-dimensional power spectrum of terrestrial gravity waves
exhibits a nearly universal behavior in its large wavenumber region
(see the previous section).  This statement is based on a large number of
observations of the vertical structure of the horizontal wind,
temperature, and density fluctuations of the present wave modes using
diverse experimental techniques.  The shape of the spectrum at large
vertical wavenumbers (short vertical wavelengths) is consistently
independent of time, place, and altitude.  This is usually attributed
to a saturation process that limits the wave amplitude growth.

Theoretical work diverges on the nature of the saturation process at
work.  The existing theories use different physical mechanisms for
dissipating wave energy including shear and convective instabilities
\citep{DewanGood1986jgrSatunivprof}, cascade processes
\citep{Dewan1991grlModigwspec}, wave-induced Doppler effects
\citep{Hines1991jasSatgwii}, and wave-induced diffusion
\citep{Gardner1994jgrDiffgwspec, Zhu1994jasSatgws}.  The different
theories all predict the same shape and behavior for the PSD,
including the -3 slope of the large wavenumber tail
and the presence of a characteristic wavenumber \math{m\sb{*}} that
limits the range of the saturated spectrum.  Since
\math{m\sb{*}\propto 1/\sigma\sb{T}}, the theories
predict a decrease in \math{m\sb{*}} with altitude as
\math{m\sb{*}\propto e\sp{-1/4H}}, and the terrestrial data agree.

Despite the diversity of wave-generating mechanisms and atmospheric
thermal structure on different planets, the saturation theories are
based on rather general concepts, implying similar spectra on
different planets.  Confirming this on other planets is challenging as
it requires local measurements at a variety of altitudes and
horizontal locations on each planet, yet there have been only a few
atmospheric entry probes.  Occultations average over a long path
length, underestimating the wave amplitudes.  This effect depends on
wavelength and probably modifies the true spectral slope, especially
at large wavenumbers.  Indeed, in Earth's atmosphere a typical
value for \math{t} is 2.5 -- 3 \citep{AllenVincent1995jgrgravwav},
whereas our best fits show a slope of 3.5 -- 4.2.  The higher values
for \math{t} might also be real, reflecting actual differences between
the dominant saturation processes acting in Earth's and Saturn's
atmospheres.  Gravity wave studies in Earth's atmosphere are based
on thousands of temperature profiles taken at different locations,
altitudes, and times of the year, using ground-based, airborne, and
satellite-based techniques, whereas we are discussing a single
temperature profile.  On the other hand, the value of
\math{L\sb{*}\approx} 15 -- 25 km, based on the observed temperature
variance (see Equation (\ref{eq:desaubies})), is in good agreement with the
value determined using the Desaubies fits (see three-parameter fits in
Table \ref{table:desaubies}).

Where should we focus our efforts for wave detection?  In addition to
seeking individual wave modes that may reveal information about
atmospheric structure before losing their identity through
interactions, one can look at the spectral characteristics at small
wavenumbers (large vertical wavelengths).  The theory says that at
wavenumbers smaller than \math{m\sb{*}}, the waves are not affected by
the saturation processes (whatever they are) and retain the spectral
characteristics of the generation mechanism.  Numerical simulations of
convectively generated gravity waves show that there is also a
dependence between the characteristic vertical wavenumber
\math{m\sb{*}} and the depth of the convective cell generating the
waves \citep{FrittsAlexander2003revgeogwdyn}.

\section{CONCLUSIONS}

We have analyzed a light curve based on IRTF observations of the 1998
November 14 occultation of GSC 0622-00345 by Saturn (Paper I).  We
presented and analyzed isothermal light-curve fits, an atmospheric
thermal profile, a wavelet analysis, gravity wave modeling, and power
spectra.  The derived thermal profile varies over 142 -- 151 K in the
pressure range 1 -- 60 {\micro}bar at a latitude of 55{\decdegree}5 S.
The vertical temperature gradient is removed by more than 0.2 K
km\sp{-1} from the adiabatic lapse rate, indicating that the
stratospheric region sounded by the occultation is statically stable.
Our thermal gradient profile shows the same alternating-rounded-spiked
appearance of other occultation profiles, including one for Saturn
\citep{CoorayEtal1998icnpsatoc}.  This shape has previously been
interpreted as evidence of gravity wave breaking
\citep{RaynaudEtal2003icjupoc, RaynaudEtal2004icjupbsco,
  YoungEtal2005icjupasi}.

Our new noise tests, based on real noise sampled from our light curve's
upper baseline rather than synthetic, uncorrelated Gaussian noise,
showed that atmospheric scintillation (and similar correlated noise
sources such as spacecraft pointing drifts) can introduce relatively
strong, spurious temperature fluctuations into the thermal profile
derived by inverting an atmospheric occultation light curve.  For our
data, the effect was many K for the raw inversions, but it mainly
appeared at longer wavelengths in power spectra.  However, the
amplitudes at the shorter wavelengths found outside the COI were still
of order 0.1 -- 1 K, comparable to the amplitudes of the gravity waves
often identified in occultation data sets.  We thus developed several
significance tests for our power-spectrum analyses, and note that
without such tests, one must be skeptical of gravity-wave detection
claims in ground-based (and possibly some space-based) occultation
inversions.

We used a wavelet analysis to search for localized gravity wave
trains.  Based on the wavelet power and the use of significance tests
only, the strongest candidate had an amplitude of 0.7 K,
\math{\lambda\sb{z}} = 12.5 km at \math{z\sb{\rm max} = -35 km}, and
lasted four cycles.  It and several shorter-wavelength features stand
over the global power spectrum at the 95% confidence level or greater.
Alternative explanations for the observed periodic structures include
sound waves, planetary waves, and non-transient features.  Without
knowing the horizontal structure of the wave we cannot rule out sound
waves or planetary waves as the cause for the temperature
fluctuations.  However, one can make the argument that, for a given
\math{\lambda\sb{z}}, planetary waves typically have lower (time)
frequencies and therefore dissipate lower in the atmosphere than the
valid region of our profile.

To take into account both phase and amplitude information, which could
improve sensitivity over our amplitude-based noise limit, we fit a
gravity-wave model to the amplitude and phase of the strongest
features in the valid region of the wavelet spectrum, following
\citet{RaynaudEtal2004icjupbsco}.  The model calculates the
temperature amplitude vs.\ height of a single, damped wave mode
propagating throughout our profile's valid region.  We performed fits
both with and without a new parameter for constant vertical wind
shear.  The added parameter did not improve the fits, so we report the
shearless fits here for consistency with prior results from this
model.

Our best candidate for a gravity wave that propagates continuously
through the valid region, as assessed by our model, had
\math{\lambda\sb{z} = 40} km, \math{\lambda\sb{h} = 200} km, and a
period of 42 minutes.  At this altitude the wave is not strongly affected
by dissipation and achieves a maximum amplitude above the observed
atmospheric region exceeding 1 K.  According to
Equation (\ref{eq:detectability}), this wave was detectable if the angle
between the line of sight and the horizontal direction of wave
propagation exceeded 74{\degrees}.  However, similar fits to
quasi-periodic features in thermal profiles derived from isothermal
(i.e., waveless) light curves with real noise gave some fits of similar
appearance.  This demonstrates the need for a study of real noise to
establish an amplitude criterion that discriminates real waves from
noise.  The reason wavelike (i.e., sinusoidal) features arise out of
the noise is simple: by reconstructing only a limited range of
wavelengths, the resulting profile is certainly sinusoidal and has a
favored period.  Wave models are sinusoidal, and ours even has
parameters that allow the phase and amplitude to vary, so we will get
a good fit if the amplitude of the reconstructed data does not vary
much.  We find that this circumstance occurs in the noise data sets
often enough to require at least criteria for significant wave
amplitude.  Criteria involving the number of cycles outside the COI or the
phase of the reconstructed data could potentially provide
even-more-stringent limits.

We derived global power spectra from our wavelet transforms, using
only data outside the COI.  This method greatly reduced the noise
level of the spectra, which stand everywhere 2 -- 10 times above the
noise level calculated using real-noise-contaminated isothermal
light curves.  The power spectra follow the modified-Desaubies form of
the universal spectrum of gravity waves, though with a slightly more
negative high-wavenumber exponent.  Superposed on this spectrum one
sees the signature of the discrete wave structures discussed above.
That we see both the universal spectrum and individual features fit
well by gravity-wave models lends confidence that we are indeed
looking at a signal dominated by gravity waves.

The amplitudes of all five wave-like features that we analyzed are
well above the noise level.  The wave model used to fit these features
is based on the assumption that the waves propagate independently of
each other.  Interactions between the wave modes might be able to explain
the discrepancies between the observed wave fluctuations and the
model's single gravity waves.  Note that the discrepancies are more
significant when derived wavenumbers are larger than the
observationally derived characteristic wavenumber.  This is the
saturated part of the spectrum where wave-wave interactions determine
wave behavior and vertical propagation.

Direct comparison of our derived temperatures with previous Saturn
occultation measurements (see, e.g., Table VII of
\citealp{HubbardEtal1997icsatmes}) requires care, both because of
differences in the assumed mean molecular mass and because Saturn's
mean stratospheric temperature is strongly affected by seasonally
varying insolation \citep{BezardGautier1985icsatclimate}.  Heating by
inertia-gravity waves might also be important, at least in
some regions and/or seasons \citep{CoorayEtal1998icnpsatoc,
FrenchGierasch1974jasocwav, YoungEtal1997scijupgravwav}.  Detailed
modeling of stratospheric temperatures requires taking account of
non-LTE effects as well \citep{Appleby1990icch4gp}.  A clearer picture
of zonal and seasonal variations in Saturn's stratospheric structure
should emerge when Earth-based stellar occultations can be viewed in
the context of data from the {\em Cassini} orbiter.

\acknowledgments

We thank L.\ A.\ Young for helpful discussions, and reviewers J.\ L.\
Elliot and E.\ Lellouch for comments that significantly improved this
paper.  We thank the NASA Astrophysics Data System, JPL Solar System
Dynamics group, and the free and open-source software communities for
software and services.  Free wavelet software
(http://paos.colorado.edu/research/wavelets/) was provided by C.\
Torrence and G.\ Compo.  This investigation was supported by Wellesley
College under NASA Contract 961169 and by the NASA PGG program.

{\em Facilities:\/} IRTF

\nojoe{
\if\submitms y
  \newcommand\bblnam{ms}
\else
  \newcommand\bblnam{satoc98-2}
\fi
\bibliography{\bblnam}}

\end{document}